\documentclass[12pt]{article}
\usepackage{graphicx}

\begin{document}
{\large\bf{ Pad\'{e}-Improvement of CP-odd Higgs Decay Rate into Two \vspace{1cm}Gluons}}
\begin{center}
F. A. Chishtie and V. Elias\\
Department of Applied Mathematics\\
University of Western Ontario\\
London, Ontario N6A 5B7, \vspace{.3cm}Canada\\
\vspace{.3cm}and\\
T. G. Steele\\
Department of Physics and Engineering Physics\\
University of Saskatchewan\\
Saskatoon, Saskatchewan S7N 5C6, \vspace{3cm}Canada 
\end{center}
\begin{abstract}
We present an asymptotic Pad\'{e}-approximant estimate 
for the four-loop coefficients within the linear combination of correlators 
entering the recently calculated decay rate of a CP-odd Higgs boson, with 
an assumed mass $m_A = 100 \; GeV$, into two gluons.  All but one of these coefficients
are shown to be determined for arbitrary $m_A$ from the known 3-loop-order rate
by renormalization group methods. Asymptotic Pad\'{e}-approximant estimates
for these coefficients are all seen to be within 12\% of their correct values. The 
four-loop term in the decay rate for $m_A = 100 \;GeV$ is estimated to be only 4.3\%
of its leading one-loop contribution.
\end{abstract}
\eject
The decay rate into two gluons $(g)$ of a CP-odd Higgs boson ($A$)
occurring within a two-Higgs-doublet extension of the standard model
has been calculated to three loop order by Chetyrkin, Kniehl, 
Steinhauser, and Bardeen \cite{Chetyrkin}.  Their result is expressed 
in terms of a linear
combination of the imaginary parts of three different correlators:
\begin{equation}
\Gamma(A \rightarrow gg) = \frac{\sqrt{2}G_F}{m_A} \,R\left(\alpha_s,
q^2 = m_A^2, \mu^2 = m_A^2, m_t^2\right),
\end{equation}
\begin{eqnarray}
R(\alpha_s, q^2, \mu^2, m_t^2) & \equiv & \tilde{C}_1^2
Im\langle [O^{\prime}_1]^2\rangle\nonumber
\\
& + & 2\tilde{C}_1 \tilde{C}_2 Im \langle
[O^{\prime}_1][O^{\prime}_2]\rangle \nonumber
\\
& + & \tilde{C}^2_2  Im \langle [O^{\prime}_2]^2\rangle . \label{Ra}
\end{eqnarray}
For $N_c = 3$ and $n$ light flavours, the terms within the linear 
combination (2) are shown \cite{Chetyrkin} to be $[x \equiv \alpha_s/\pi, L \equiv \ln(\mu^2/q^2),
L^{\prime} \equiv \ln(m^2_t/q^2)]$
\begin{equation}
\tilde{C}_1 = \frac{-x}{16} \left[1 + {\cal{O}}(x^3)\right],
\end{equation}
\begin{equation}
\tilde{C}_2 = x^2 \left[\frac{1}{8} - (L -
L^{\prime})/4\right] + {\cal{O}}(x^3),
\end{equation}
\begin{eqnarray}
Im \langle [O^{\prime}_1]^2\rangle & = & \frac{8q^4}{\pi} \left\lbrace 1 +
x\left[\left(\frac{97}{4} - \frac{7n}{6}\right) + \left(\frac{11}{2} -
\frac{n}{3}\right)L\right]\right.\nonumber
\\
& + & x^2 \left[392.223 - 48.0753n + 0.887881n^2\right.\nonumber
\\
& + & \left( \frac{3405}{16} - \frac{73}{3}n + \frac{7}{12}
n^2\right)L\nonumber
\\
 & + &  \left.\left. \left(\frac{363}{16} - \frac{11}{4}n +
\frac{n^2}{12}\right) L^2\right] + {\cal{O}}(x^3)\right\rbrace ,
\end{eqnarray}
\begin{equation}
Im \langle [O^{\prime}_1][O^{\prime}_2]\rangle  = \frac{q^4 xn}{\pi} + {\cal{O}}
(x^2),
\end{equation}
\begin{equation}
Im \langle [O^{\prime}_2]^2\rangle  = \frac{q^4 x^2n^2}{8\pi} + {\cal{O}}
(x^3).
\end{equation}
One can combine these results to obtain the following 
series for the linear combination of correlators defined by (2):
$$
R(\alpha_s, q^2, \mu^2, m_t^2)  \equiv  \frac{q^4}{32\pi} S[x, L,
L^{\prime}]\nonumber$$
$$ = \left(\frac{q^4}{32\pi}\right) x^2 \left[1 + (a_0 + a_1 L) x +
\left(b_0 + b_1 L + b_2 L^2\right)x^2\right.\nonumber$$
$$\left.+ \left(c_0 + c_1 L + c_2 L^2 + c_3 L^3\right)x^3 +
\ldots\right], \eqno(8a)$$
$$a_0  =  \frac{97}{4} - \frac{7n}{6}, \eqno(8b)$$
$$ a_1  =  \frac{11}{2} - \frac{n}{3}, \eqno(8c)$$
$$b_0  =  392.223 - (48.5753 + L^{\prime})n +
0.887881n^2,\eqno(8d)$$
$$b_1  =  \frac{3405}{16} - \frac{70n}{3} + \frac{7}{12} n^2
,\eqno(8e)$$
$$b_2  =  \frac{363}{16} - \frac{11n}{4} + \frac{n^2}{12} .\eqno(8f)$$

The terms listed above arise entirely from the first two terms of (2), 
as the final term $(\tilde{C}_2^2 Im\langle [O_2^{\prime}]^2\rangle )$
is ${\cal{O}}(x^6)$.  The four-loop $[{\cal{O}}(x^5)]$ coefficients 
$c_0 - c_3$ in (8) are as 
yet undetermined. All but $c_0$ of these can be obtained via renormalization-group 
(RG) methods.  RG-invariance of the physical decay rate (1) and, consequently, the linear 
combination of correlators (2) implies that the function $S[x,L,L^{\prime}]$ in (8a) satisfies
$$\left[ \frac{\partial}{\partial L} + \beta (x) \frac{\partial}{\partial x} + 2\gamma_{m_t} (x) 
\frac{\partial}{\partial L^{\prime}}\right] S[x, L, L^{\prime}] = 0,
\eqno(9)$$
where
$$\beta(x) = - \beta_0 x^2 - \beta_1 x^3 -\beta_2 x^4 \ldots
,\eqno(10)$$
$$\gamma_{m_t} (x) = -\gamma_0 x - \gamma_1 x^2 - \gamma_2 x^3 \ldots \;\;\;
.\eqno(11)$$
If $m_t$ is a pole-mass independent of the renormalization scale $\mu$,
then $\gamma_{m_t} = 0$.  However, if $m_t$ is a $\mu$-dependent running quark mass,
then $\gamma_0 = 1$, and subsequent $\gamma_i$'s in (11) are as determined in ref. \cite{Chetyr}.

Substitution of (8a), (10), and (11) into (9) yields the following set of equations 
for the aggregate coefficient of $x^{k_{1}}L^{k_{2}}$ to vanish:
$$\hspace{-1.4cm}x^3\!\!: a_1 - 2\beta_0 = 0, \eqno(12)$$
$$\hspace{-2.6cm}x^4\!\!: b_1 - 3\beta_0 a_0 - 2\beta_1 = 0, \eqno(13)$$
$$\hspace{-3.3cm}x^4 L\!\!: 2b_2 - 3a_1\beta_0 = 0,\eqno(14)$$
$$\hspace{.2cm}x^5\!\!: c_1 - 4b_0 \beta_0 - 3a_0 \beta_1 - 2\beta_2 + 2n\gamma_0 = 0,
\eqno(15)$$
$$\hspace{-1.8cm}x^5 L\!\!: 2c_2 - 4b_1 \beta_0 - 3a_1 \beta_1 = 0, \eqno(16)$$
$$\hspace{-3cm}x^5 L^2\!\!: 3c_3 - 4b_2 \beta_0 = 0.\eqno(17)$$
The coefficients $\beta_{0-2}$ in (10) for 
$n$ light flavours are given by \cite{Caso}
$$\beta_0 = \frac{11}{4} - \frac{n}{6},\eqno(18)$$
$$\beta_1 = \frac{51}{8} - \frac{19n}{24}, \eqno(19)$$
$$\beta_2 = \frac{2857}{128} - \frac{5033n}{1152} +
\frac{325n^2}{3456};\eqno(20)$$
the coefficients $\gamma_1, \gamma_2, \ldots$
in (11) do not enter (9) until ${\cal{O}}(x^6)$.
Using eqs. (8.b,c,e,f), (18), and (19), we see that eqs. (12-14) are explicitly upheld, 
thereby confirming the RG invariance of (8a). The unknown coefficients $c_1$, 
$c_2$, and $c_3$ are 
obtained via equations (15-17):
$$c_1 = 4822.88 - n(11L^{\prime} + 884.455 + 2\gamma_0) + n^2\left(\frac{2}{3}
L^{\prime} + 45.1091\right) - 0.591921 n^3,\eqno(21)$$
$$c_2 = \left( \frac{11}{2} - \frac{n}{3} \right) \left(\frac{1779}{8} -
\frac{1177n}{48} + \frac{7n^2}{12} \right),\eqno(22)$$
$$c_3 = \frac{1}{2} \left(\frac{11}{2} - \frac{n}{3}
\right)^3.\eqno(23)$$
For the physical case of $n = 5$, $m_t = 175.6 \; GeV$ [the t-quark pole mass $(\gamma_0 = 0)]$,
with $m_A$ chosen as in [1] 
to have a reference value 
of $100 \; GeV$, we find that 
$$c_1 = 1411,\quad c_2 = 438.4,\quad c_3 = 28.16\;\;.\eqno(24)$$
The coefficient $c_0$ is RG-inaccessible to order $x^5$.

The four-loop correlation-function coefficients $c_{0-3}$ can be estimated using 
asymptotic Pad\'{e}-approximant methods as delineated in ref. \cite{Chishtie}. 
Given a correlation function of the form
$$\Pi(x) = F(x) \left[1 + R_1 x + R_2 x^2 + R_3 x^3 + \ldots \right],
\eqno(25)$$
with only coefficients $R_1$ and $R_2$ known, the simplified asymptotic error formula 
(utilized in 
\cite{Ellis} to estimate $\beta_3$ from $\beta_{0-2}$)
$$\delta_{N+2} \equiv \frac{R_{N+2}^{[N|1]} - R_{N+2}}{R_{N+2}} = \frac{-
A}{N+1}, \eqno(26)$$
characterizing the $[N|1]$ Pad\'{e}-approximant prediction for 
$R_{N+2}$, yields the following prediction
for $R_3$ \cite{Elias}:
$$R_3 = 2R_2^3/(R_1^3 + R_1 R_2).\eqno(27)$$
Comparing eq. (25) to (8a), we see that the coefficients $R_1$, $R_2$, $R_3$ are necessarily 
functions of $L = \ln(\mu^2/q^2)$:
$$R_1 = a_0 + a_1 L,\eqno(28a)$$
$$R_2 = b_0 + b_1 L + b_2 L^2,\eqno(28b)$$
$$R_3 = c_0 + c_1 L + c_2 L^2 + c_3 L^3.\eqno(28c)$$
Consequently, we can obtain $c_{0-3}$ from the moment integrals 
$$N_k \equiv (k+2) \int_0^1 \,dw\, w^{k+1} R_3(w),\eqno(29)$$
where $w \equiv q^2/\mu^2 [L = -\ln(w)]$.\footnote{Such integrals characterize the $O(x^3)$
contributions to the $k^{th}$ finite-energy sum rule integral
$\int_0^{s_0} t^k \Pi(x,t)dt$ over the correlator (25), where $q^2$ in (29) corresponds
to $t$, and where $\mu^2$
in (29) corresponds to the continuum threshold $s_0$.} Explicit substitution of (28c) 
into (29) yields \cite{Chishtie}
$$N_{-1} = c_0 + c_1 + 2c_2 + 6c_3,\eqno(30)$$
$$N_0 = c_0 + \frac{1}{2} c_1 + \frac{1}{2} c_2 + \frac{3}{4}
c_3,\eqno(31)$$
$$N_1 = c_0 + \frac{1}{3} c_1 + \frac{2}{9} c_2 + \frac{2}{9}
c_3,\eqno(32)$$
$$N_2 = c_0 + \frac{1}{4} c_1 + \frac{1}{8} c_2 + \frac{3}{32}
c_3.\eqno(33)$$
Numerical values of $N_{-1}$, $N_0$, $N_1$, and $N_2$ can be obtained 
through explicit use of 
the Pad\'{e}-motivated estimate (27) within the integrand of (29) with $R_1$ and $R_2$ given 
by (28a) and (28b).  Within these latter two equations, the coefficients $a_0$, $a_1$, $b_0$,
$b_1$, $b_2$ are as given by (8.b-f).  We choose $n = 5$ and $L^{\prime} = 2\; {\rm{ln}} (1.756)$ 
to facilitate 
comparison with the true RG values (24) for $m_A = 100\; GeV$, and find that 
$$N_{-1} = 3310.4,\quad N_{0} = 1863.8,\quad
N_{1} = 1512.6,\quad N_2 = 1359.2\; .\eqno(34)$$
We substitute these values into (30-33) to find that
$$c_0 = 981.7,\quad c_1 = 1274, \quad c_2 = 452.3, \quad c_3 = 24.96\;\;
.\eqno(35)$$
The relative errors of the above Pad\'{e} estimates for $c_1$, $c_2$, and $c_3$ from their true 
values, as given in (24), are respectively -9.7\%, +3.2\%, and -11.3\%.

\begin{figure}[hbt]
\centering
\includegraphics[scale=0.7]{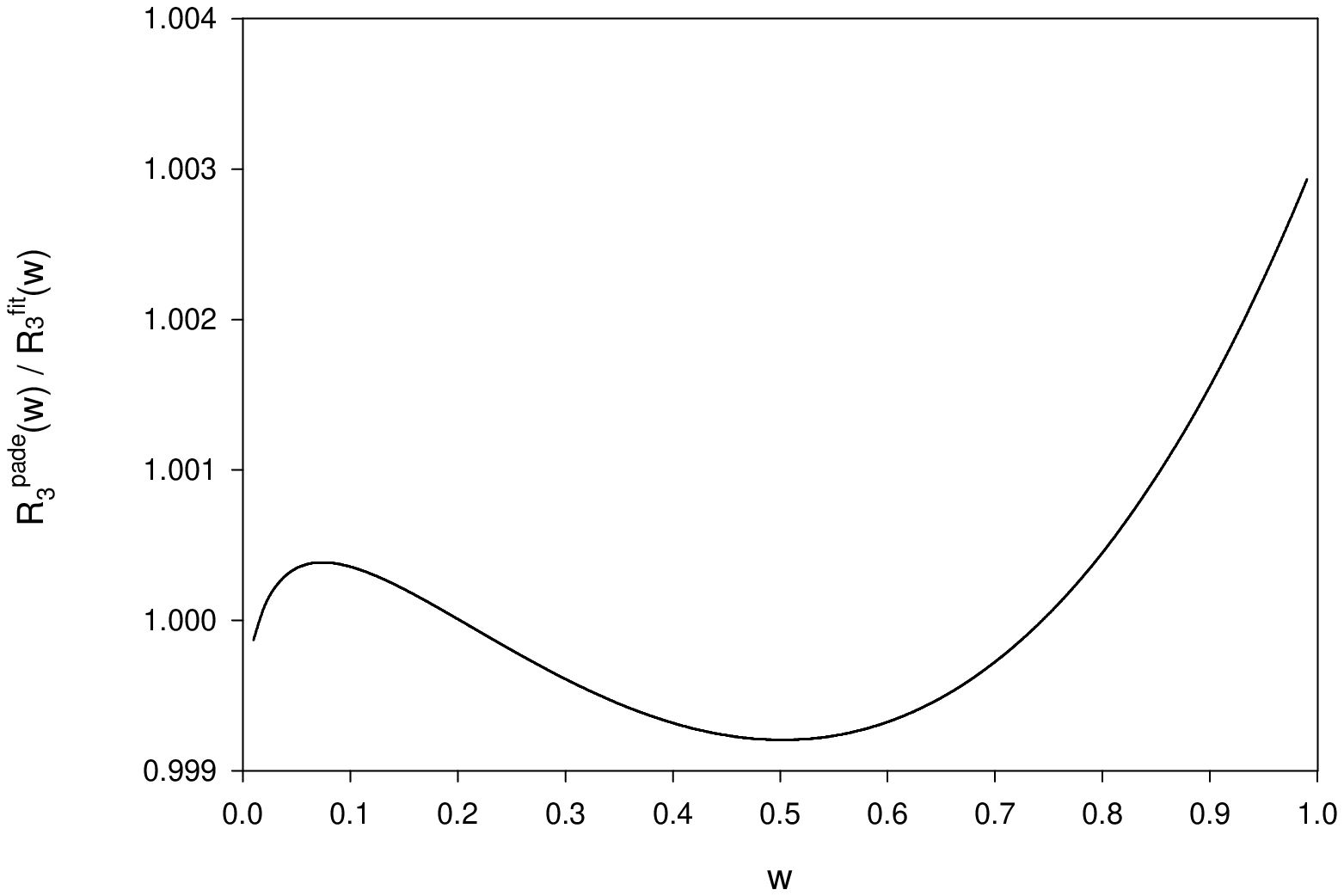}
\caption{The ratio of the Pad\'e prediction  $R_3^{pade}(w)$ (27), and the $\chi^2$-minimizing
fitted form  $R_3^{fit}(w)$ (28c)
as a function of $w$.}
\label{chi2_fig}
\end{figure}

An alternative method for extracting $c_{0-3}$ is to fit $R_3(w)$, as obtained
from (27), to the form of (28c) via least-squares minimization of the following function:
$$\chi^2 (c_0, c_1, c_2, c_3) 
=\int_0^1 [R_3(w) - (c_0-c_1 ln(w) + c_2 ln^2(w) - c_3 ln^3(w))]^2 dw \nonumber$$
$$=0.2532\cdot 10^8 + 720 c_3^2 + 12 c_0 c_3 + 24c_2^2 \nonumber $$
$$+48 c_1c_3 + 2c_1^2 + 240 c_2 c_3 + c_0^2 + 12 c_1 c_2 \nonumber $$
$$+2 c_1 c_0 + 4 c_2 c_0 - 6620.83 c_0 - 13688.4 c_1 \nonumber $$
$$-46945.6 c_2 - 217719 c_3. \eqno(36)$$
The values of $c_{0-3}$ which minimize $\chi^2$ are 
$$c_0 = 978.9, c_1 = 1285, c_2 = 445.0, c_3 = 26.03, \eqno(37)$$
in very close agreement with the values (35) extracted from the moment integrals (29).
Moreover, $\chi^2$ is equal to only 4.7 at this minimum.  The near cancellation of the $O(10^7)$ lead term
in (36) is indicative of the precision of the fit obtained between (27) and (28c), as is evident from
Figure \ref{chi2_fig}.  Relative errors of $\{c_1, c_2, c_3\}$ obtained from (36) with respect to their true
RG-determined values (24) are respectively -8.9\%, +1.5\%, and -7.6\%, confirming the usefulness
of the asymptotic Pad\'e approach in estimating four-loop order contributions to the correlation
function (25).  

The $m_A = 100\; GeV$ estimate for $c_0$ can be improved somewhat by using the 
correct values (24) of $c_{1-3}$ within equation (30), the lowest moment integral 
estimated in (34) by asymptotic Pad\'{e}-approximant methods.  We then find that
$$c_0 = 3310.4 - 1411 -2(438.4) - 6(28.16) = 854.\eqno(38)$$
Identically the same result is obtained by minimizing $\chi^2$ with respect to $c_0$ after explicit
incorporation of the correct (RG) values (24) of $c_{1-3}$ into (36).
There is a 15\% discrepancy between  (38) and the estimates for $c_0$ in (35) and (37), 
indicative of the magnitude of anticipated relative error with respect to the
true value for $c_0$.  It is hoped that these estimates can be tested against an exact 4-loop
calculation in the not-too-distant future.

The 4-loop correction to the CP-odd Higgs decay into two gluons, as
determined to 3-loop order in \cite{Chetyrkin}, is found from (1), (8)
and (38):
$$\Gamma(A \rightarrow gg) = \frac{\sqrt{2} G_F m_A^3}{32\pi} \left[1 +
a_0 x(m_A) + b_0 x^2(m_A) + c_0 x^3(m_A)\right].\eqno(39)$$
Given $n = 5$, $m_A = 100\; GeV$, $\alpha^{(5)}(m_A) = 0.116$
\cite{Chetyrkin}, and $m_t = 175.6 \; GeV$, the square bracketed expression in
(39) for successive-loop corrections is $[1 + 0.680 + 0.226 +
{\underline{0.043}}]$. The first three numbers are as calculated in ref.
\cite{Chetyrkin}; the final (underlined) term is obtained from the
asymptotic Pad\'{e}-approximant estimate for $c_0$ in (38).  
This estimate is further indicative of a progressive decrease in the ratio of 
successive terms in the $A \rightarrow 2g$ decay rate, suggesting that if such
a CP-odd Higgs were discovered, a perturbative calculation of its 2-gluon decay
rate could lead to a phenomenologically testable value
[{\it e.g.} 1 + 0.680 + 0.226 + 0.043 + ... $\approx$ 2.0 for $m_A$ = 100 GeV].  Such a 
Higgs characterizes the two-doublet version of electroweak symmetry breaking anticipated 
from supersymmetric extensions of the standard model, as first noted over two
decades ago \cite{Sherry}. 

\bigskip

{\large\bf{Acknowledgment}}
Support from the Natural Sciences and Engineering Research Council of Canada is 
gratefully acknowledged.

\end{document}